\documentclass[aps,letterpaper,amssymb,twocolumn,pra]{revtex4}
\usepackage{amsmath}
\usepackage{amssymb}

\newcommand{\nn}{\nonumber\\}
\usepackage{bm}
\usepackage{amssymb}
\usepackage{times}
\usepackage{latexsym}
\usepackage{graphicx}
\usepackage{color}
\usepackage{subfigure}
\begin{document}
\title{ Supermetallic and Trapped States in Periodically Kicked Lattices}
\author{ Indubala I Satija and Bala Sundaram  }
\affiliation{Department of Physics and Astronomy, George Mason University,
 Fairfax, VA 22030}
\affiliation{         	 
Department of Physics, University of Massachusetts at Boston,100 Morrissey Blvd
Boston, MA 02125}
\date{\today}
\begin{abstract}
A periodically driven lattice with two commensurate spatial periodicities is found to exhibit {\it super metallic} states
characterized by enhancements in wave packet spreading and entropy. These resonances occur at critical values of parameters
where multi-band dispersion curves
reduce to a universal function that is topologically  a circle
and the effective quantum dynamics describes free propagation.
Sandwiching every resonant state are a pair of anti-resonant {\it trapped states} distinguished by dips in entropy where the
transport, as seen in the spreading rate, is only somewhat inhibited.  Existing in gapless phases fo the spectrum, a sequence of these peaks and dips are interspersed by
gapped phases assocated with {\it flat band states} 
where both the wave packet spreading as well as the entropy exhibit local minima.
\end{abstract}
\maketitle

Periodically driven systems exhibiting novel topological properties
have been the subject of numerous recent studies, motivated by new 
methods to achieve and control topological structures by external driving~\cite{kicked,kicked1}.
In addition  to $Z_n$ invariants, the Chern numbers, non-equilibrium systems exhibit nontrivial topology that require new invariants such as
the winding number of the quasienergy. Topological aspects characterized by the winding numbers
have no analog in the corresponding static systems and their importance has emerged in several recent studies on topological insulators,
seen as exotic states of matter that are insulating in the bulk but conduct along the edges~\cite{TIreview}.
In particular, winding numbers of the quasienergy have been associated with topologically protected edge modes~\cite{kicked1}, in a
manner unrelated to the Chern numbers.
 
In contrast to previous studies that focus
on insulating states of matter with topologically protected gapless edge modes, the central focus of this paper is on metallic phases exhibiting ballistic transport,
characterized by the quadratic spreading of wave packets in time. We find that
at special parameter values, where the quasienergy bands are associated with integer winding numbers, metallic states are found to exhibit
strong enhancement in the transport properties. We refer to this behavior as {\it super metallic} and show that these are resonances
where the multi-band structure collapses to a single band with a dispersion curve resembling a Lissajous figure.
The resonances for different windings are described by a universal function that gives rise to an effective Hamiltonian
exhibiting free-particle dynamics. 
These Lissajous resonant states are accompanied by another type of behavior
characterized by a sharp drop in the entropy associated with an evolving wavepacket though there is little corresponding effect on
the spreading. This latter phenomenon is accompanied by a phase jump of $\pi$ at neighboring sites near the localization center,
suggestive of destructive interference.
It should be noted that both behaviors occur only
when the spectrum is gapless corresponding to metallic transport at all temperatures.

Though we stress situations where the spectrum is gapless, we also find a special set of parameter values
corresponding to gapped phases that exist inbetween the metallic regions. At these critical values transport as measured by wave packet 
spreading exhibits local minima as does the entropy. These points are found to correlate with novel band structure as the quasienergy bands
exhibit a partial flattening and hence correspond to zero group velocity.  Overall, the system under investigation exhibits extremely rich and 
complex band structure that leads to a number of interesting dynamical effects.

The periodically kicked spatially modulated lattice considered is described by the following Hamiltonian.
\begin{eqnarray*}
    H& =& \frac{1}{2}J\sum_m \left(\hat{b}^{\dag}_m \hat{b}_{m+1}+ \textrm{H.c.}\right)\\
&+&K \sum_m \cos{(2\pi\sigma m)}\hat{b}^{\dag}_m \hat{b}_{m}
 \sum_n \delta(t/\tau-n),
\label{Hamil},
\end{eqnarray*}
where $\hat{b}^{\dag}$ is the creation operator for the quantum particle, be it bosonic or a fermionic. Here $J$ is the nearest-neighbor hopping
term and $K$ describes the strength of the stroboscopic time dependent onsite potential with period $\tau$.
The parameter $\sigma$ introduces a competing periodicity and with the choice
$\sigma=1/q$, where $q$ is an integer, the resulting superlattice is periodic with a period $q$. In the time-independent situation where the onsite
potential is always on, periodic lattices are characterized by the band structure and Bloch states, namely the eigenstates and eigenvalues of the
Hamiltonian. Analogously, periodically driven systems are studied in terms of the  properties of  
their time-evolution operators $U(\tau)$ acting over one full period of the drive.  
Each eigenstate of $U$, called a Floquet state, accumulates a phase over one period of 
the driving which is the quasienergy associated with that state. 
However, because the quasienergy is defined as a phase variable, it is periodic with period $2\pi$. 
This periodicity introduces a topological structure, associated with the winding of quasienergy, which has no analog in static systems. 
 
Since our Hamiltonian is periodic in time, we have
\begin{equation}\label{floquet}
    \psi(t)\equiv \psi_\omega(t)=e^{-i\omega t}\phi_\omega(t),
\end{equation}
with $\phi_\omega(t+\tau)=\phi_\omega(t)$. The non-equilibrium system is described by the eigenvalue equation,
\begin{equation}\label{ev eq}
    \hat{U}\psi_\omega(t)=e^{-i\omega\tau}\psi_\omega(t).
\end{equation}
where $\omega$ is the quasienergy of the system, while $\phi_\omega$ is the corresponding quasienergy state.
For systems with discrete translational symmetry, we index the quasienergies by the crystal momentum $k$
and define a quasienergy band structure. The phase factors or quasienergies may exceed a value of
$2\pi$ leading to
the winding of the quasienergy bands as the Bloch index $\kappa$ runs over the first Brillouin zone (BZ).
For simplicity, we will set $\tau=1$ and therefore, the quasienergies are measured in the unit of the driving frequency $\tau^{-1}$.

For the special case of $\sigma=1/2$, the system maps onto a spin-$1/2$ system and can be solved analytically.
The quasienergies are given by,
\begin{equation}
\omega_{1,2} (\kappa)= \pm Arcos[\cos(\bar{K}) \cos (\bar{K} \zeta)]
\label{U2}
\end{equation}
where $\bar{K}=\frac{K}{2\pi\sigma}$ and $\zeta=\cos(\pi \kappa)$.
The corresponding eigenvectors are,\\
\begin{align}
    \psi_{\omega_1} &= \left[\cos\beta ,\; \sin\beta e^{i\theta}\right], \nn
    \psi_{\omega_2} &= \left[\sin\beta ,\; \cos\beta e^{i\theta}\right],
\label{U2wf}
\end{align}
where $\beta$ and $\theta$ are given by
$\tan\beta=\sin(\bar{K})\cot(\bar{K}\zeta)+\sqrt{1-\cos^2(\bar{K})\cos^2(\bar{K}\zeta)}/(\sin(\bar{K}\zeta))$
and $\theta=(\bar{K}+\kappa)$.

With the exception of this exactly solvable $\sigma=1/2$ case, lattices for all other values of $\sigma$ 
require numerical diagonalization of a matrix representation of $U$ to obtain their spectral features.
In addition to spectral properties, we also consider the temporal dynamics 
of an initial state. We have investigated a variety of initial conditions including initial states localized at a single or
multiple sites as well as Gaussian wave packets. We considered
lattices of various sizes upto $2^{18}$, varying also the total number of
kicks, where longer time evolutions require bigger lattice sizes in order to avoid boundary effects.
 
We note that due to the spatial periodicity of the lattice, all quasienergy states are extended. Therefore,
irrespective of the value of the parameter $K$, the  wave packet grows ballistically in time, 
with root mean square displacement increasing quadratically, $\sum_m |\psi_m|^2 m^2 \propto t^2$.
In order to quantify the temporal dynamics we construct scaled variables defined as,

\begin{eqnarray}
X^2& =& \frac{1}{T^2} \sum_m |\psi_m|^2 m^2\\
S &= & \frac{1}{T}e^{ - \sum_m |\psi_m|^2 ln |\psi_m|^2},
\end{eqnarray}
where $T$ is the total number of kicks that an initial condition is subjected to.

\begin{figure}
\includegraphics[width=.65\textwidth, height=.65\textwidth]{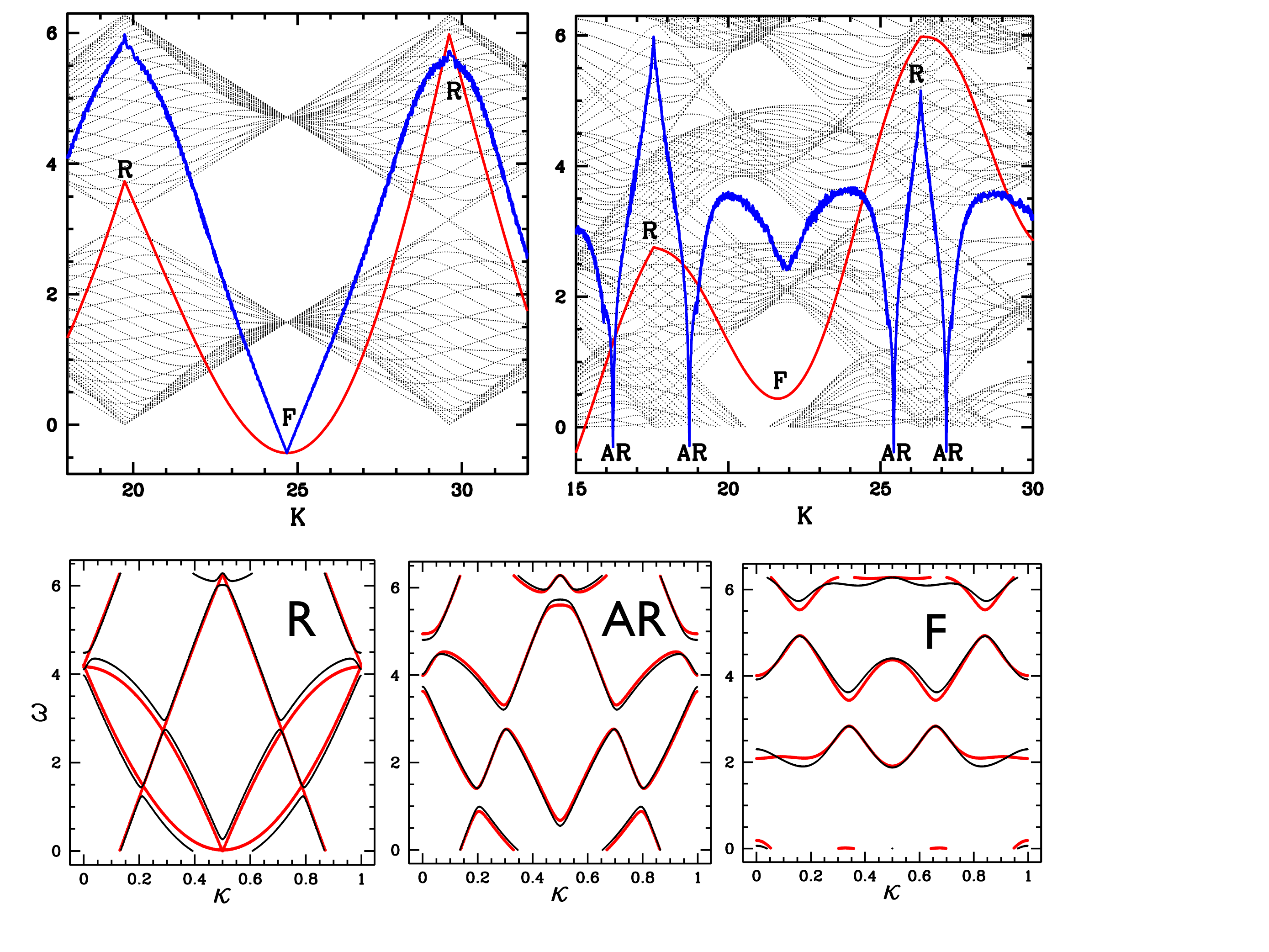}
\leavevmode \caption {(color online) Upper panels show quasi-Energy spectrum (black), and the wave packet spreading
 $\bar{X^2}$ (red) and the entropy $S$ (blue) for $\sigma=1/2$ ( left ) and $\sigma=1/3$ (right),
as a function of $K$ on a lattice with $2^{15}$ sites and $T=3000$. The special values of $K$ denoted by R, AR and F are discussed in the text.
The bottom panel shows the quasienergy dispersions, for $\sigma=1/3$ at (red) and near (black) these special points. Note that at values corresponding
to R the distinct bands join smoothly while at the F values the bands are quite flat. By contract the spectrum at AR shows
no clearly discernible characteristics.}
\label{Eall}
\end{figure}

Figure~\ref{Eall} illustrates changes in spectral and temporal characteristics as the kicking parameter $K$ is varied. 
The quasienergy spectrum in the bulk, for both values of $\sigma$, shows gapless as well as gapped phases with considerably
more spectral complexity seen in the case of $\sigma=1/3$. 
As indicated in the figure, we identify three distinct types of behaviors, referred as super-metallic or Lissajous resonant (R),
a contrasting anti-resonant (AR) and finally flat-band (F), where the reasoning behind this nomenclature will soon be made clear. In terms
of the transport measures, the R states are distinguished by local maxima in $X^2$ and $S$ while AR states display
sharp dips in $S$. There is no clear evidence of AR in $X^2$. By contrast, the F states which appear in the gapped phase exhibit 
local minima in both $X^2$ and $S$. By scanning the kicking parameter $K$, a cascade 
of parametric windows hosting these states at special values are observed. The simple $\sigma=1/2$ case does not exhibit
AR behavior and, in our numerical study of various commensurability parameters, $\sigma=1/3$ is the simplest case that captures the
key characteristics and, as such, we will use it to illustrate our findings.

\begin{figure}%
\centering
\subfigure{\includegraphics[width=1.5in,height=1.5in]{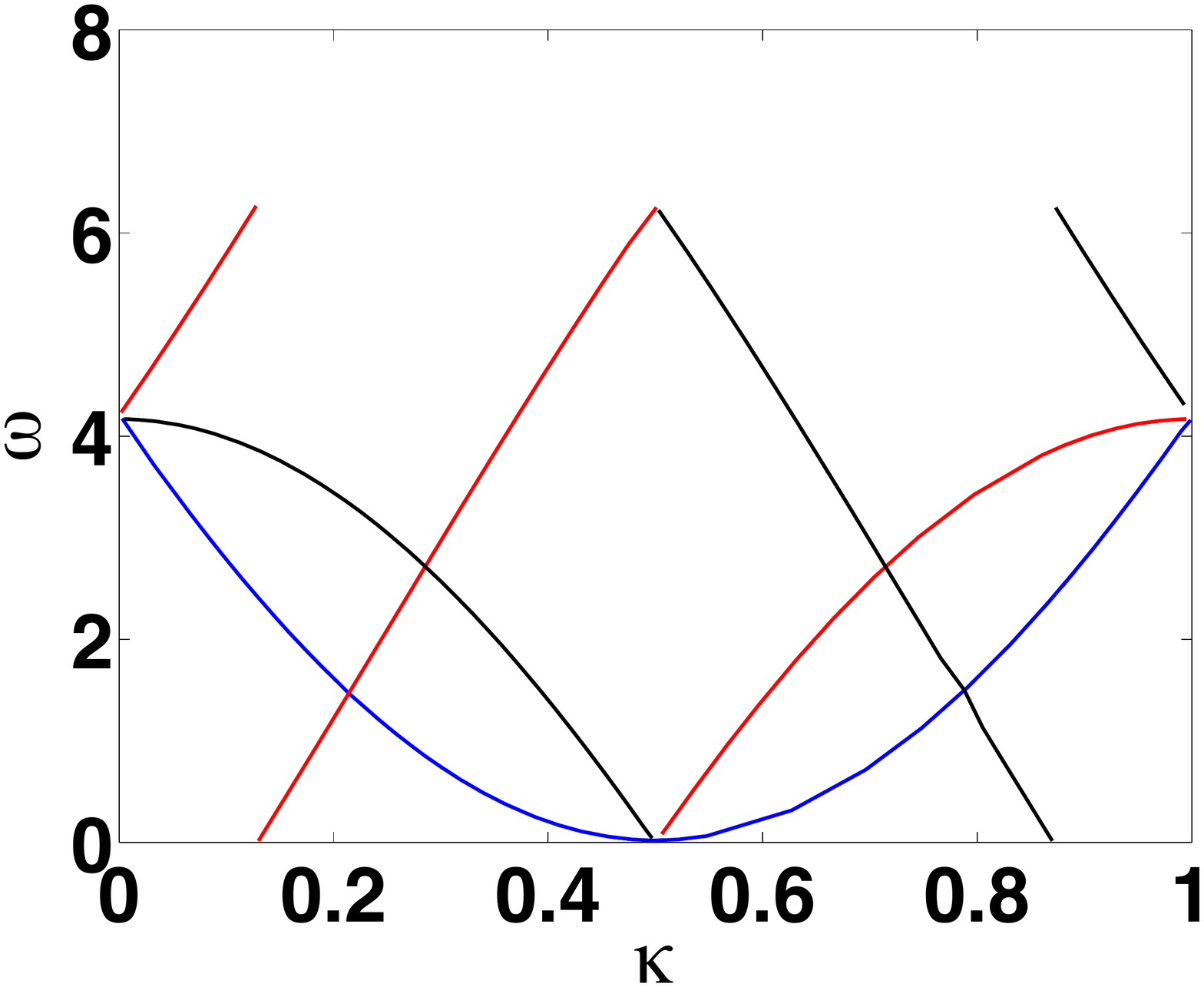}} \label{SingleCa}
\subfigure{\includegraphics[width=1.5in,height=1.5in]{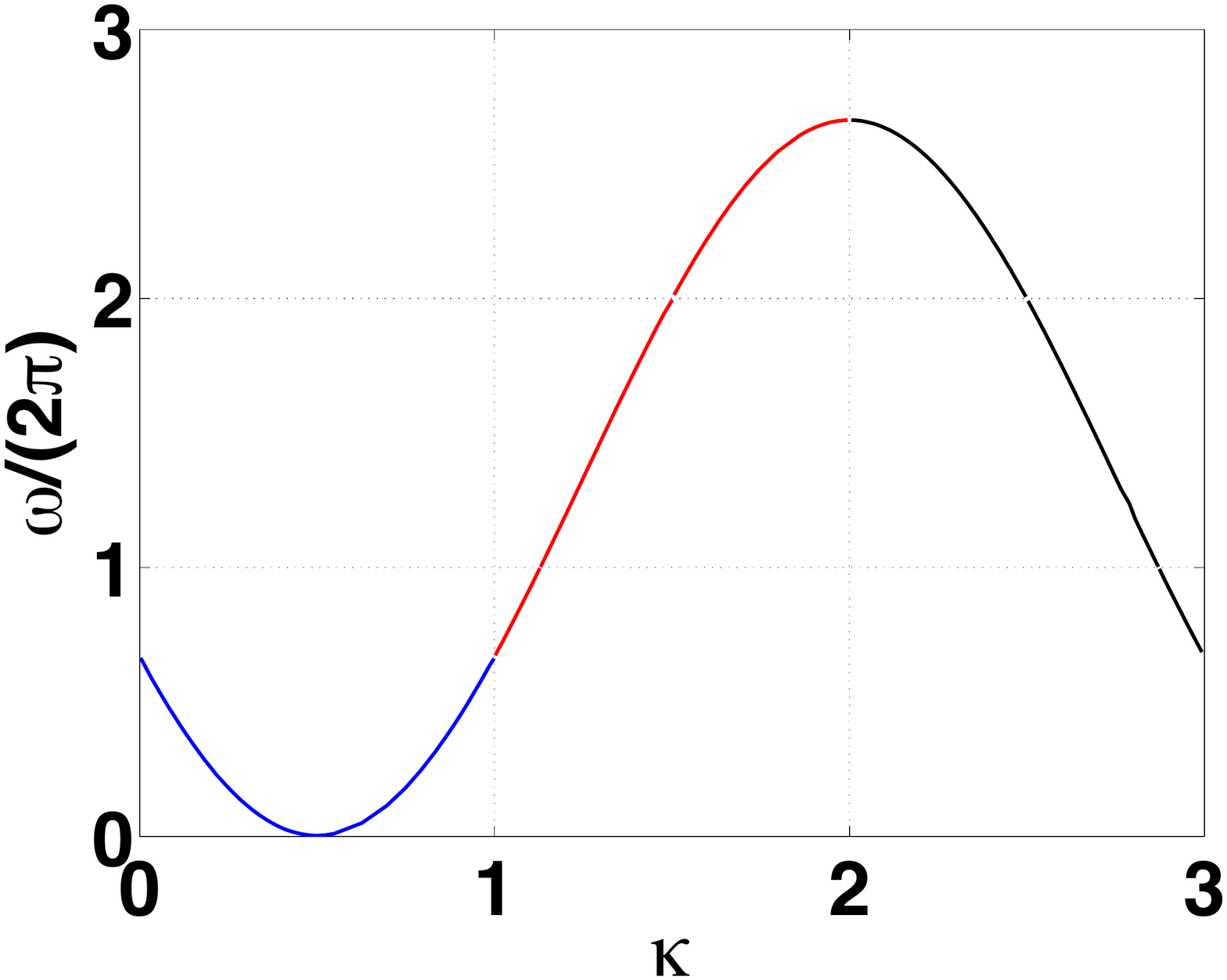}} \label{SingleCb} \\
\vspace{-0.3in}
\subfigure{\includegraphics[width=2.9in,height=2.6in]{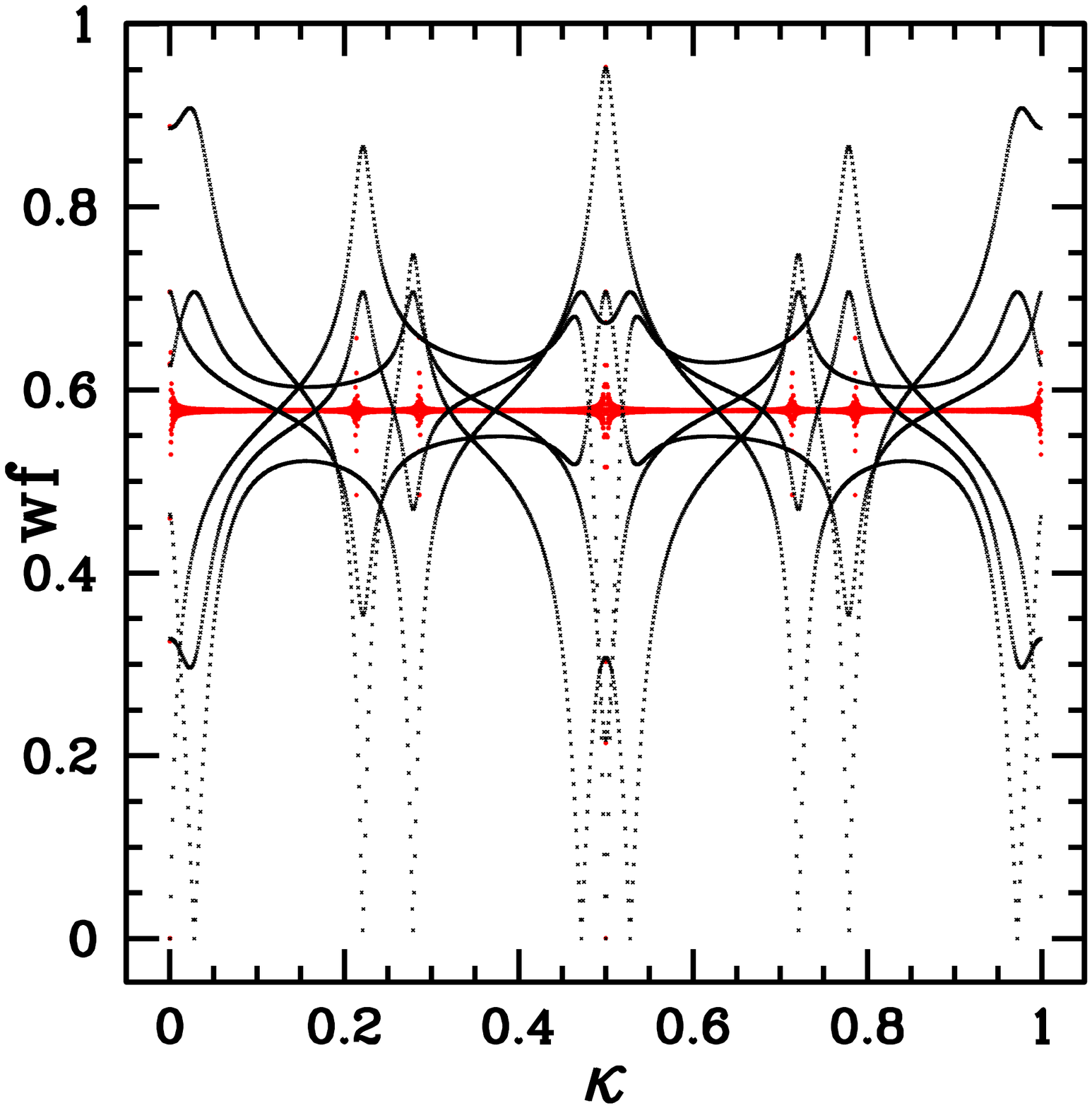}}\label{SingleCc}
\vspace{-0.5in}
\caption {(color online) (Top row) The two figures illustrate the collapse of the multi-band structure into a single band at $K_2^*$. (Left) the three bands are
shown while in (Right) their representation as a dispersion curve in the extended Brillouin zone is shown. This occurs at all the locations
exhibiting  resonant (R) behavior seen in Figure ~\ref{Eall}. We note that periodicity of $\omega$ by multiples of $2\pi n$ along with the periodicity
of the BZ is essential in this construction and the figure is topologically a circle. As seen from the figure,
with the net winding number equal to zero, the extended BZ consisting of three BZ can be associated with winding numbers $0$, $2$ and $-2$.
The lower panel shows the corresponding Bloch states as a function of Bloch index $\kappa$ at ( in red) and near ( in black) the AR 
behavior. The AR Bloch states have equal amplitudes for all values of $\kappa$ except where the bands intersect.}
\label{SingleC}
\end{figure}

We begin with a detailed analysis of the super metallic resonant states which occur for parameter values resulting in a gapless spectrum.
We note that the band structure for arbitrary $\sigma=1/q$ consists of a family of $q$-dispersion curves that overlap in energy. 
In a periodically driven system, where the quasienergy is defined as a phase factor
and may change by an integer multiple of $2\pi$ as one traverses the BZ, the possibility emerges that the quasienergy
bands could meet (actually intersect tangentially) at critical values of the parameter governing the spectrum ($K$). In view of the 
periodicity of both the quasienergy and the reciprocal space,
for integer windings of the quasienergy, a union of q-dispersion curves results in a band structure that resembles a Lissajous figure as illustrated in the Fig.~\ref{SingleC}.
Furthermore, the Lissajous figure seen in the fundamental domain of the BZ
can be mapped to a single curve, also shown in Fig.~\ref{SingleC}, that is not only continuous but also
an analytic function of the Bloch index $\kappa$ in an extended BZ of size $q$ times the size of the original BZ.
Away from these special $K$ values, this intriguing union of dispersion curves does not occur as seen in Fig. ~\ref{Eall} by
contrasting the band structure at and near R. 

For $\sigma=1/3$, the Lissajous resonances occur at $K^*_n = 2n (2\pi\sigma)^2$ with the band structure for different $n$-values described, in the extended
BZ, is found to fit into a single universal function, 
\begin{equation}
\frac{\omega^*(\kappa)}{2\pi} = n(1-\sigma)\left[ 1 -\cos 2\pi \sigma( \kappa-\kappa_0) \right]\; ,
\end{equation}
with $\kappa_0=\frac{1}{2}$. As illustrated in Fig. \ref{SingleC} ( upper panel), the band in the extended BZ  with net winding number zero can be identified with three winding numbers, $(0, n, -n)$
corresponding to the winding numbers in each of the three regions of BZ that form the extended zone. Furthermore, the band is topologically a circle, suggestive of a metallic state that
may be topologically nontrivial.

The spectral function $\omega^*(\kappa)$ can be viewed as resulting from an effective Hamiltonian
of a free particle, resulting in free propagation of matter waves. Taking into account the fact that $X^2$ exhibits a peak as well at these $K$
values indicating maximal transport, we will refer to these resonant states as super metallic. This nomenclature is supported on viewing the
Bloch states describing the super metallic states which have equal amplitudes at all sites, that is independent of the Bloch index
$\kappa$ ( see lower panel in Fig. ~\ref{SingleC}), except at isolated points
where the curves intersect. The resonant state can be viewed as a kind of {\it ergodic} state
as all allowed quasienergy states are uniformly populated, differing only in their phases that assume all possible values.
The existence of these Lissajous bands is a unique feature
of driven lattices with no counterpart in the equivalent static systems where energy has an upper and lower bounds.  
We note that the resonant behavior characterized by the functional form $\cos(2\pi\sigma \kappa)$ appears to be a generic feature of the system valid for all values of $\sigma$
as verified numerically for $\sigma=1/4, 1/5, 1/6$ and by analytic calculation for $\sigma=1/2$ shown below.

\begin{figure}
\includegraphics[width=.55\textwidth]{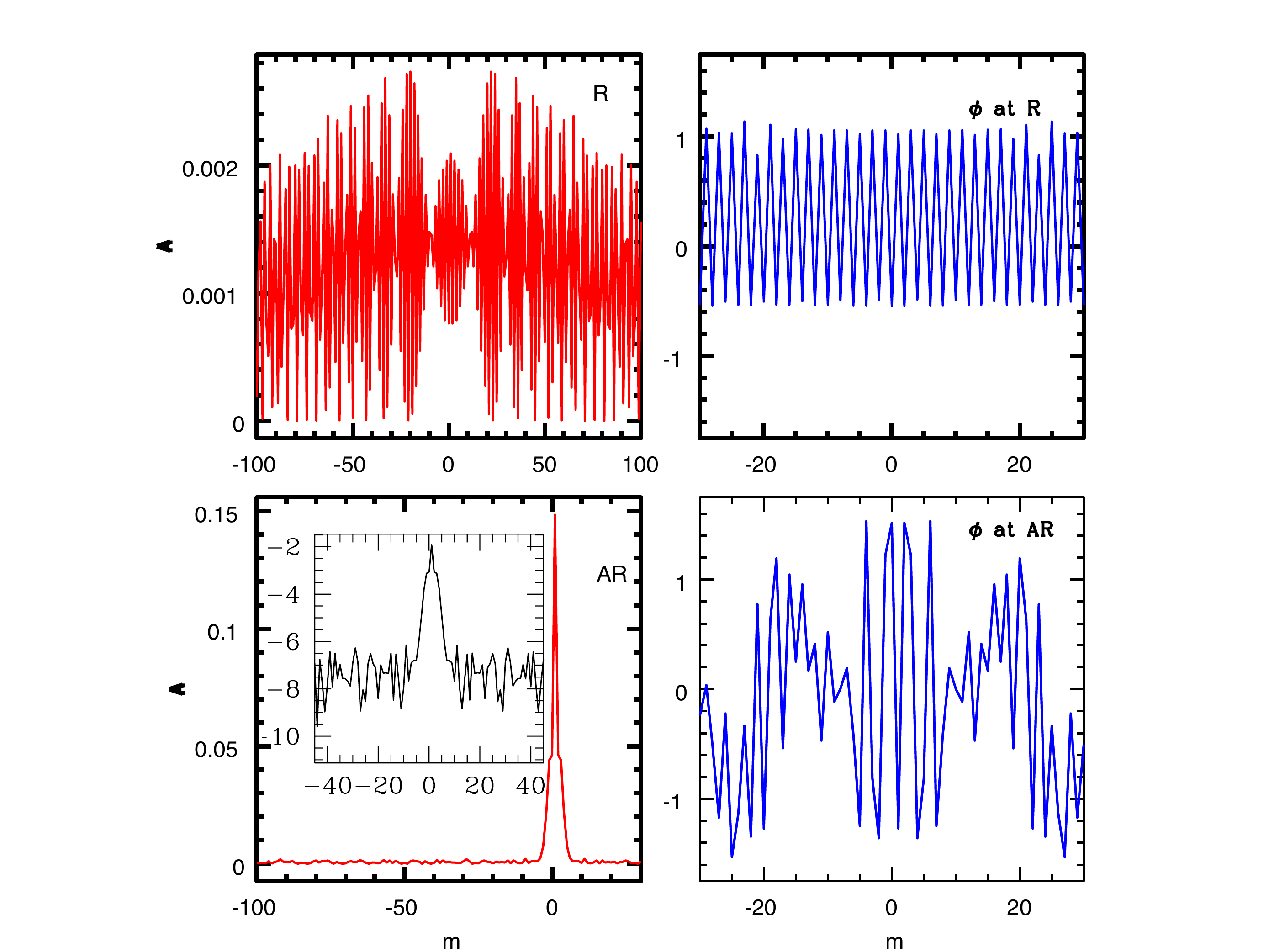}
\leavevmode \caption {(color online)
Spatial profiles of the amplitudes (left) and phases (right), of  a time evolved wave packet. 
In all cases,  an initial Gaussian wave packet with width $\sqrt{2}$ was time evolved for $T=500$ at parameters corresponding to the
Lissajous resonance (top) and antiresonance (bottom).
The insert in the bottom panel shows the profile on a semi-logarithmic scale,
illustrating the coexistence of localized and extended character at the anti-resonances. In terms of the phase variations
from lattice point to the next, this phase is distinguished by phase jumps of $\pi$ between neighboring sites which are near the initial site.}
\label{wf3d}
\end{figure}

The integrable case of $\sigma=1/2$, that exhibits resonances at  semimetallic diabolic points~\cite{diabolic} which occur at
$K^*= n (2\pi\sigma)^2$, provides a simple analytic illustration of the various characteristics associated with the resonant behavior described.
At these critical values, the quasienergy dispersion (\ref{U2}) simplifies to
$\frac{\omega_{\pm}}{2\pi} = \pm n(1-\sigma)(1 +\cos 2\pi \sigma \kappa)$.
In the extended BZ, these two curves can be described by a single curve
$\frac{\omega^*}{2\pi} =2 n(1-\sigma) cos 2\pi \sigma\kappa$.
Furthermore, as seen from Eq. (\ref{U2wf}),
the magnitude of the quasienergy wave functions becomes $\kappa$ independent and all states are
equally populated with $\beta=\frac{\pi}{4}$ and $\theta= \pm \kappa$ illustrating ergodic character of the resonances mentioned above. 

An intriguing aspect of the quantum dynamics in the metallic gapless phase is
the presence of characteristics dips in the entropy, occurring in close proximity to the resonant points.
These are in fact satellite dips that accompany every resonance as revealed in our studies with different values of $\sigma$.
We emphasize that these dips occur inside the parametric windows where the bands overlap. However, the dips are not associated with
any special spectral features of the bulk and the wave packet spreading remains relatively
immune to these sharp decreases in the entropy as illustrated in Fig. (\ref{Eall}).
Further, as seen from Fig. ~(\ref{wf3d}), a time evolved initial condition shows distinct characteristics both in terms of the
amplitude at lattice sites as well as in the phase of the projections. The spatial profile of the time evolved state displays both localized
and extended components. It should be noted that the wave packet remains localized at the initial site for all finite times, with the height of the peak decreasing very slowly with increasing $T$. 
The phase of the wave packet projected onto lattice sites shows a clear phase jump of $\pi$ for neighboring sites near the
localization center.  This is in sharp contrast to what is seen at the Lissajous resonances, where the wavepacket is seen to be
quite clearly extended and projections show phase jumps of $\pi/2$ between neighboring sites. This
difference in phase projections motivated our characterization of the dips as manifestations of anti-resonant behavior. 
As this anti-resonant phenomena also occurs in the gapless phase, in close proximity to the resonant points, these satellite dips
are absent in cases where the resonant point coincides with a diabolic point  ( semi-metallic phase ) as seen for
$\sigma=1/2$ and numerous other $\sigma$ values such as $1/4$ and $1/6$.  We also note here that band overlaps, of the type shown in the
lower-middle panel of Fig.~\ref{Eall}, appear necessary for these dips to occur. Characteristic phase differences of $\pi$ between neighboring sites
associated with these coherent anti-resonances is reminiscent of Fano-type resonances~\cite{Fano} although a detailed exploration of quasienergy states with open boundary conditions did not reveal the coexistence of a discrete state and a continuum of states necessary for the Fano mechanism.
Therefore, the presence of sharp drop in entropy at two points sandwiching every resonance point  is dynamically generated effect who origin remains somewhat mysterious.

Returning to Fig.~\ref{Eall}, we notice another feature which we will now address. The series of peaks in $X^2$ are separated by local minima
in the wave packet spreading, again at special parameter values except now these are in the gapped phase.
Close examination of the band structure at these points reveal partially flattened band structure as seen in the F spectrum shown in Fig.~\ref{Eall}.
We note that these points have analogies with the integrable case of $\sigma=1/2$ where
the minima occur at $K = \frac{n}{2} \pi^2$ where the quasienergy band collapses to $\omega_{\pm} = \pm \pi/2$ for
all values of the Bloch index $\kappa$. 
Critical parameters corresponding to local minima in $X^2$ are also accompanied by a local minima in the entropy.
At these points, as expected from these features, the temporal evolution of an initial condition exhibits strong localization at the starting location.

Arising from the integer windings of the quasienergies, 
and characterized by a universal band structure that is topologically a circle,
the super metallic states are
robust against various perturbations
due to their insensitivity to continuous deformations of the quasienergy spectrum.
This suggests that these resonant states may describe metallic states that are topologically nontrivial.
With no counterpart in the corresponding static system, this phenomena results
from the commensurability of quasienergies, the phase factors accumulated during quantum evolution,
and the periodicity of the reciprocal space. 
We would like to reiterate that the super-metallic behavior described here refers to enhancement in the transport that is ballistic,
as root mean square displacement grows quadratically with time, and hence is quite distinct from other phenomena
referred to as super diffusive. Also the possible topological aspect is unrelated to the states that have been referred as
topological-metallic states~\cite{TM}. Further, the dynamically generated anti-resonant states which accompany this behavior
are particularly intriguing and deeper understanding of their existence
remains somewhat elusive. 

Finding new topological states of matter continues to remain
an active frontier in condensed matter and AMO physics.
With intense research focused on topological insulators,
our results suggest new avenues involving states that
are metallic and whose origin in tied to invariants that describe global properties of quasienergy states.
These resonances suggest new ways to engineer band structures that leads to states of enhanced conductivity.
Similarly, the associated anti-resonances
that correspond to almost localized wave packet may provide a new method to create trapped states.

Superlattices have been studied in ultracold~\cite{twocolor} and also photonic laboratories~\cite{photonic}.
Furthermore,  periodically driven systems have been realized in laboratories as Floquet topological insulators\cite{FloquetTI} and also in a context 
related to quantum chaos~\cite{Qchaos}. 
We note that the resonances observed here may have counterparts in the semiclassical limit
and hence may be analogous to transitions described in earlier studies~\cite{SSP}.
Finally, periodically driven superlattices may serve as new paradigms to explore recent emergent phenomena
seen in chains that are open to the environment~\cite{Prosen} and may simulate new avenues of exploration in periodically driven quantum Hall~\cite{ML} as well as in superconducting chains that host Majorana modes~\cite{SC}. 

We would like to thank Erhai Zhao for his help, suggestions and comments during the course of this work.
This research is supported by ONR (IIS) .

\end{document}